\documentclass[14pt,a4paper,twoside,fleqn]{extarticle}
\usepackage{els-chem-crc}
\usepackage{times}
\usepackage{amssymb,cite}

\usepackage{graphicx}
\usepackage[figuresright]{rotating}
\configuresideways

\newcommand{\AmS}{{\protect\the\textfont2
  A\kern-.1667em\lower.5ex\hbox{M}\kern-.125emS}}
\newcommand{\figwidth}{0.95\columnwidth}

\chapter{1}
\title{Electronic transport and localization in short and long DNA}

\author{H. Wang, R. Marsh, J. P. Lewis,
\address[Lewis]{Department of Physics and Astronomy, Brigham Young University,
Provo, UT 84602-4658, U.S.A.}
        R.\ A.\ R\"{o}mer\address[Roemer]{Department of Physics and Centre for Scientific Computing, University of Warwick, Coventry CV4 7AL, United Kingdom}
        \thanks{Work partially supported by the Royal Society.}
        }

\begin{document}
\maketitle

\begin{abstract}
  The question of whether DNA conducts electric charges is intriguing to
  physicists and biologists alike. The suggestion that electron
  transfer/transport in DNA might be biologically important has
  triggered a series of experimental and theoretical investigations.
  Here, we review recent theoretical progress by concentrating on
  quantum-chemical, molecular dynamics-based approaches to short DNA
  strands and physics-motivated tight-binding transport studies of long
  or even complete DNA sequences. In both cases, we observe small, but
  significant differences between specific DNA sequences such as
  periodic repetitions and aperiodic sequences of AT bases,
  $\lambda$-DNA, centromeric DNA, promoter sequences as well as
  random-ATGC DNA. %
  ($Revision: 1.15 $)\newline%
\end{abstract}

\section{Introduction}
\label{sec-introduction}

Charge transfer in DNA is currently the subject of intense theoretical
and experimental investigations
\cite{Schuster2362004,Schuster2372004,Singh2004,PorCD04}. DNA, which is
the blueprint of life, is being considered as a molecular wire in a new
generation of electronic devices and computers. However its electronic
properties are elusive and remain controversial. Despite the current
debate, the subject is far from new. Soon after Watson and Crick
discovered the double-helix structure of DNA \cite{Crick1953}, Eley and
Spivey were the first to suggest that DNA could serve as an electronic
conductor \cite{Spivey1962}. The notion of a molecular wire is thought
to apply to the DNA double helix because of its $\pi$- electron (the
$\pi$- way) system of bases stacked upon each other.  More recently,
Barton and colleagues \cite{KelB99} measured the fluorescence of an
excited molecule and found that it no longer emitted light when attached
to DNA. Their results suggested that this ``fluorescence
quenching\char`\"{} was due to the charge on the excited donor molecule
leaking along the length of the DNA to a nearby acceptor molecule.

Other extensive experimental and theoretical work over the past decade
has led to substantial clarification of charge-transfer mechanisms in
DNA
\cite{KelB99,Michel-Beyerle1999,Schuster1999,Zewail1999,Giese1201998,Giese371998,
  Tanaka1998,Wasielewski1997,Harriman1994,Ratner2001,Schuster2001,Michel-Beyerle1998,
  Risser1997,Beratan1996}.  The dominant mechanisms appear to be both
short-range quantum mechanical tunneling and long-range thermally
activated hopping. Guanine has the highest occupied molecular orbital
(HOMO) level of the four bases, and can act as a trap for holes.
Experiments on repeats of this base are used to investigate long range
hopping, and models have been developed to clarify the long range
hopping data in G-repeats \cite{Wasielewski1997}.  Charge transport in
DNA is also made more complex because of the influences of the local
environment, such as counterions, thermal vibrations, contact
resistance, and sequence variability, which are difficult to control
\cite{Anantram2003,Ratner2002,Artacho2000,Frauenheim2002,CunCPD02}. The
charge-transfer mechanisms in DNA and/or whether DNA is a good conductor
or not remains somewhat unsettled. Indeed, theory is of great help in
understanding these phenomena, but given the computational cost of
full-scale calculations on the realistic DNA systems, theoretical
efforts to date have mostly been limited to small- and medium-size model
systems \cite{Steenken2002,Saito1996,Yoshioka1999}, to dry DNA molecules
\cite{Artacho2000,Lewis2004,Sankey2003}, or to larger systems using
model Hamiltonians
\cite{Roc03,RocBMK03,Igu03,RocM04,YamSHA04,Yam04,Yam04b,Igu04,Shi05a,Shi05b,KloRT04,KloRT05}
and semi-empirical studies
\cite{Ratner2002,Rudnik2000,Ducasse2003,Beratan2002,
  Rosch2001,Sta05,HenSAP04,Sta02,Sta02b,Sta04b,TanNSS04,CueSAH04}.

In this review, we shall first focus on the use of quantum-chemical
methods which can treat smaller, but atomistically correct segments of
DNA in Sec. \ref{sec-short}.  After an introduction to the construction
of the DNA molecules and the density-functional based methods in Sec.
\ref{sec-short-generating_structures} - \ref{sec-short-localization}, we
then present results, many of which are new, in sections
\ref{sec-short-periodicDNA} to \ref{sec-short-dynamicDNA}. In the next
large section \ref{sec-long}, we use the lessons learned from the
atomistic approach and now study an effective and necessarily rather
coarse-grained Hamiltonian model of DNA to reveal the interplay of
sequence fidelity and transport. Again, models, methods and DNA
sequences are introduced in Sec. \ref{sec-long-ladder} -
\ref{sec-long-DNA}. Sections \ref{sec-long-results} and
\ref{sec-long-promoters} include the obtained results.  We conclude and
summarise in Sec. \ref{sec-summary}.

\section{Quantum chemical methods for {\it short} DNA strands}
\label{sec-short}


Within a density functional based local orbital tight-binding-like
formalism, more complex problems can be investigated with a modest
decrease in the accuracy. This is particularly useful where a quantum
mechanical description is important to the investigated system's
fundamental chemistry, yet where a smaller model system would
inadequately describe the proper physical environment. With the increase
in computational power, great effort has been made by the
electronic-structure community to optimise the performance of quantum
mechanical methods. Calculating larger systems without making stringent
approximations has only been possible within the past few years. In this
chapter, we theoretically investigate the electronic states of model DNA
structures as the molecule undergoes classical thermal motion at room
temperature by means of marrying classical molecular dynamics
simulations with an electronic structure density-functional method. We
investigate the dynamics of the DNA structure and its impact on the
electronic structure.  A similar approach was recently used to postulate
the charge migration mechanism in DNA, with injected charges being gated
in a concerted manner by thermal motions of hydrated counterions
\cite{Schuster2001}. Here we study a longer oligonucleotide duplex than
previous studies, and demonstrate with the complete system that its
electronic states dynamically localize.  The mechanism is an Anderson
\emph{off-diagonal} dynamic disorder model similar to the static
disorder that leads to localised band-tail states in amorphous
semiconductors \cite{Anderson1958,Drabold1998,Overhof2001,Gotze1981}.
The concept of static Anderson localization in DNA has previously been
considered by Ladik \cite{Bakhshi1986,Ladik2001}.  We show that
localization in DNA reaches far deeper in energy than just band tail
states. We demonstrate for the first time this effect in a hydrated
poly(dA)-poly(dT) 10 base-pair fragment; this represents one complete
turn of the B-DNA double helix.

\subsection{Generating the poly(dA)-poly(dT) DNA structures}
\label{sec-short-generating_structures}

In this chapter, we consider thermal fluctuations of a poly(dA)-poly(dT)
DNA 10-mer duplex fragment at room temperature from classical MD
simulations; therefore, aperiodic structures of DNA are generated
throughout the simulation. With our local-orbital density-functional
method, we compare the electronic states of an idealised model periodic
canonical B-DNA poly(dA)-poly(dT) DNA structure with those
thermally-distorted aperiodic poly(dA)-poly(dT) DNA structures generated
from the MD simulation.

Canonical B-DNA 10-base pair models of poly(dA)-poly(dT) were built into
a Arnott B-DNA \cite{Hukins1972} model using the nucgen DNA builder
contained within AMBER 5.0 \cite{Kollman911995}. Classical molecular
dynamics trajectories of the B-DNA models, including explicit water and
sodium counterions, were generated using the CHARMM (version c26n1)
\cite{Karplus1983}.  Both models were solvated with enough
pre-equilibrated TIPSP \cite{Karplus1983} water to add 12.0 \AA~to the
maximal distance extent of the DNA.  Net-neutralising Na+ ions
\cite{Aqvist1990} were placed off the phosphate oxygen bisector and then
minimised (with larger, 5.0 , van der Waals radii) in-vacuo prior to
solvating the system. Equilibration involved the application of harmonic
positional restraints (25.0 kcal/mol$^2$) and 250 steps of ABNR
minimisation, followed by 25 ps of MD where the temperature was ramped
up from 50 to 300 K in 1 ps intervals.  The initial equilibration was
performed with the Cornell et al.\ force field \cite{Kollman1171995}.
Subsequent equilibration with the ??? (BMS) force field of Langley
\cite{Langley1998}, involved 250 steps of ABNR minimisation followed by
5 ps of MD with position restraints.

All production simulations were performed without any restraints and the
BMS force field of Langley. Production simulation was performed for 10
nanoseconds with CHARMM (version c26n1) \cite{Karplus1983} in a
consistent manner. This involved constant temperature (300 K, mass =
1000) \cite{Hoover1985} and pressure (1 atm, piston mass = 500 amu,
relaxation time = 20 ps$^{-1}$) \cite{Brooks1995}, 2 fs time steps with
the application of SHAKE \cite{Berendsen1977} on hydrogen atoms,
accurate use of the particle mesh Ewald method \cite{Pedersen1995}
(\textasciitilde{}1.0 grid size with $6^{th}$ order B-spline
interpolation and a Ewald coefficient of 0.34) in rhombic dodecahedral
unit cells ($x=y=z$, $\alpha=60^{\circ}$, $\beta=90^{\circ}$,
$\gamma=60^{\circ}$), a heuristically updated atom based pairlist built
to 12.0 \AA~and cutoff at 10.0 \AA~with a smooth shift of the van der
Waals energies.  These methods have proven reliable for representing DNA
duplex structure \cite{McConnell2000,Kollman2000} and the BMS force
field very accurately models B-DNA crystal structures
\cite{Langley1998,Young2001}.

\begin{figure}[tb]

\begin{center}\includegraphics[%
  width=\figwidth]{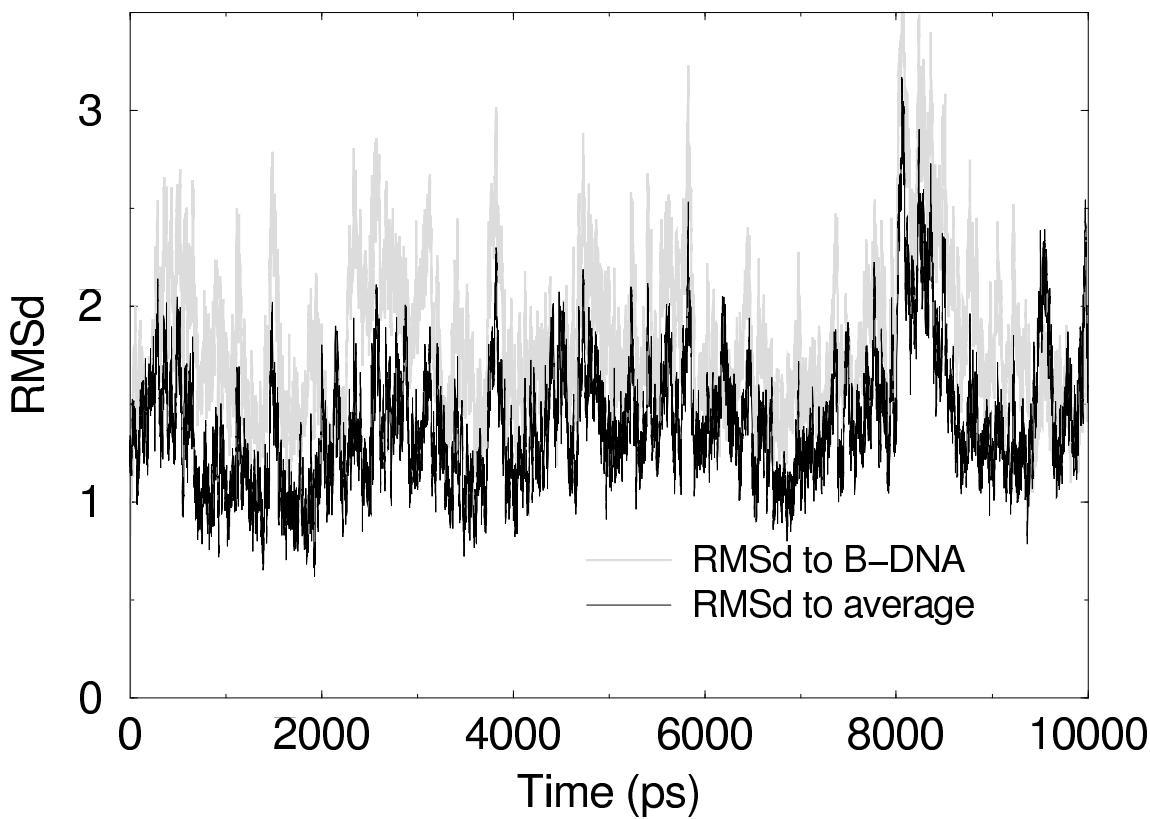}\end{center}
\caption{\label{figrmsd}
  Shown in black and gray are the all-atom best-fit root-mean-squared
  deviations (in \AA) as a function of time compared to canonical B-DNA
  (gray) and the straight coordinate average structure from the 1.5-2.5
  ns portion of the trajectory (at 0.5 ps intervals).  }
\end{figure}

After an initial equilibration of an explicitly solvated 10-mer B-DNA
poly(dA)-poly(dT) with explicit Na+ ions, production molecular dynamics
simulations (applying an accurate particle mesh Ewald treatment of the
electrostatics) were performed for 10 ns. As shown in Fig.\ 
\ref{figrmsd}, a plot of the all-atom root-mean-squared deviation over
the entire run is rather stable, and although thermal fluctuations are
clearly evident, no large scale distortions of the structure were
observed (beyond sugar repuckering, and expected base and backbone
fluctuations).

\subsection{Electronic structure calculations of molecular dynamics
snapshots}
\label{theory-fireball}

A stable portion of the trajectory from 1.5-2.5 ns, at 0.5 ps intervals,
was analysed further using the F{\small IREBALL} DFT methodology
\cite{Sankey2001}.  F{\small IREBALL} is a first principles
tight-binding molecular dynamics (TBMD) simulation technique based on a
self-consistent version of the Harris-Foulkes
\cite{Harris1985,Haydock1989} functional \cite{Demkov951618}. In this
method, confined atomic-like orbitals are used as a basis set for the
determination of the occupied eigenvalues and eigenvectors of the
one-electron Hamiltonian. The ``fireball" orbitals, introduced by Sankey
and Niklewski \cite{Sankey893979}, are obtained by solving the atomic
problem with the boundary condition that the atomic orbitals vanish
outside and at a predetermined radius $r_c$ where wavefunctions are set
to be zero. This boundary condition is equivalent to an ``atom in the
box" and has the effect of raising the electronic energy levels due to
confinement. An important advantage of the Sankey and Niklewski basis
set is that the Hamiltonian and the overlap matrix elements of the
system are quite sparse for large systems, reducing overall computation
time.  A summary of the method is given in Ref. \cite{Sankey2001} and
references therein.  All poly(dA)-poly(dT) DNA atoms, including
phosphate groups and backbone atoms are included in the single-point
calculations which contained 10 base pairs (644 atoms). Although the MD
simulations are performed with full hydration and counterions, we
include only 350 water molecules in our electronic structure
calculations; this number of molecules represents approximately 2
solvation layers surrounding the molecule. Adding all water and cation
atoms to more correctly represent the environment surrounding the DNA
molecule will be the subject of future work.

\subsection{Quantifying the Degree of Localization}
\label{sec-short-localization}

The phenomena of Anderson localization \cite{Anderson1958,RomS03} refers
to the localization of mobile quantum mechanical entities, such as spin
or electrons, due to impurities, spin diffusion, or randomness. Anderson
localization applied to DNA may come from two distinct mechanisms,
\emph{diagonal} or \emph{off-diagonal} disorder\emph{. Diagonal}
disorder induced localization occurs from variations of the sequence
along the base stack, and \emph{off-diagonal} disorder occurs by
variations either from bonding between bases along the stack or from
hydrogen bonding variations across the double helix. The qualitative
physics of localization is described by an Anderson model
\cite{Anderson1958},
\begin{equation}
H={\sum_{i}}\epsilon_{i}c_{i}^{\dagger}c_{i}+\sum_{i,j}^{'}t_{i,j}c_{i}^{\dagger}c_{j}+t_{j,i}c_{j}^{\dagger}c_{i},\label{eq:anderson}
\end{equation}
where each molecular orbital (MO) $i$ of a base has energy
$\epsilon_{i}$ and interacts with its nearest neighbour base MO $j$
($i\neq j$) with a Hamiltonian hopping interaction of t$_{ij}$. The
Anderson model of \emph{diagonal} disorder randomly varies the on-site
Hamiltonian matrix elements (diagonal) $\epsilon_{i}$ \cite{RomS03} and
describes the A-T-G-C random sequencing of DNA \cite{Roc03}.

Here we focus on B-DNA structures of poly(dA)-poly(dT) in which there
exists only one base pair combination A-T; each strand has only a single
type of base in its stack. In this system, only \emph{off-diagonal}
disorder \cite{BisCRS00} may occur. The bonds within a single base are
strong, but thermal fluctuations coupled with weak $\pi$-bonding occurs
along the stack and the weak hydrogen bonds across the strands of the
DNA double helix allows individual bases significant freedom of
movement, including transient base pair opening and DNA breathing events
over millisecond time scales \cite{Fritzche1999} and large fluctuations
in the structure \cite{Hock1998}. Stochastic fluctuations of the weak
bonding modulates the electronic coupling, $t_{ij}$, between adjacent
bases. If the dynamic fluctuations of $t_{ij}$ are large enough,
localised electronic states are produced as in an amorphous solid.

We quantify the spatial extent of an electronic state by defining the
number of accessible atoms, $W$, from the electronic state quantum
\emph{entropy}. From a particular state $\nu$, the wavefunction
$\psi(\nu)$ has a Mulliken population $p_{i}(\nu)$ on atom $i$, which
loosely is considered the probability that an electron in state $\nu$
and resides on a particular atom $i$. The populations are normalised,
$\sum_{i}p_{i}(\nu)=1$. From probability theory, we define a quantum
entropy for state $\nu$ as, \[ S(\nu)=-\sum_{i}p_{i}(\nu)\ln
p_{i}(\nu).\] For example, a state $\nu$ with equal probabilities over
N$_{0}$ atoms ($N_{0}\leq N_{Total})$, gives an entropy of $\ln N_{0}$.
{}From Boltzmann's equation, we can determine the number of accessible
atoms $W(\nu)$ for electronic state $\nu$ as $S(\nu)=\ln W(\nu)$, or
\[ W(\nu)=e^{S(\nu)}.\] Our example state with equal probabilities
spread over $N_{0}$ atoms gives the expected result, $W(\nu)=N_{0}$. For
the complex electronic states of DNA, the number of accessible atoms
$W(\nu)$ gives a quantitative, and easily calculable, measure for how
many atoms a particular electronic state $\psi(\nu)$ reaches.

\subsection{Electronic states of a periodic poly(dA)-poly(dT) DNA}
\label{sec-short-periodicDNA}

\begin{figure}[tbhp]

\begin{center}\includegraphics[%
  width=0.90\columnwidth]{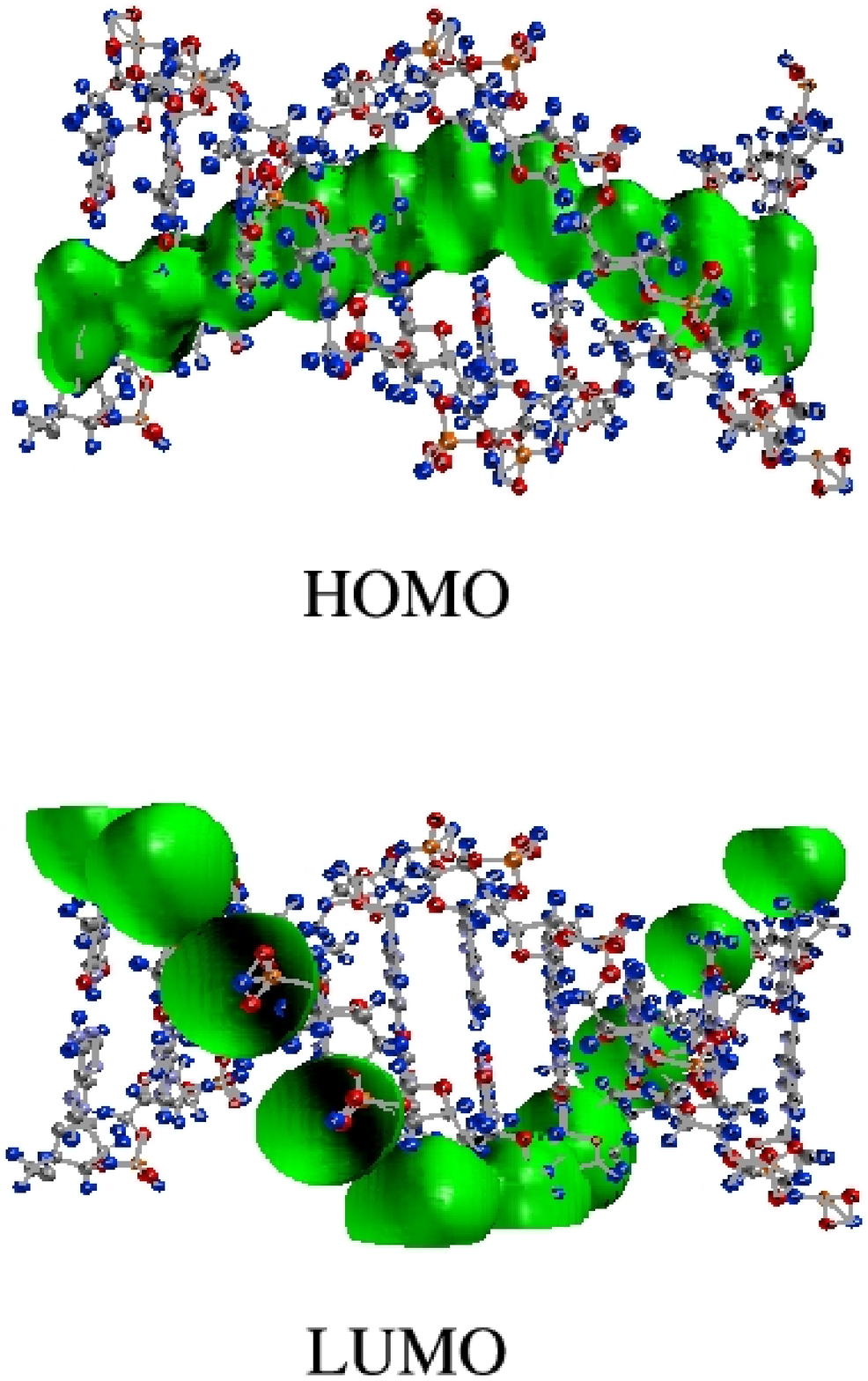}\end{center}
\caption{\label{figHLperiodic}
  Population densities for the highest occupied molecular orbital (HOMO)
  and the lowest unoccupied molecular orbital (LUMO) are shown for
  periodic poly(dA)-poly(dT) DNA (10 base pairs).  Both molecular
  orbitals exhibit very extended and periodic (Bloch-like)
  states throughout the molecule.
}
\end{figure}

To demonstrate that localization is not due to limitations of using
localized orbitals, a 10-base pair periodic structure of
poly(dA)-poly(dT) was created based on the Arnott B-DNA
\cite{Hukins1972} fiber model. Each base pair is rotated by $36^{\circ}$
and translated by $3.38$ \AA; therefore, 10 base pairs complete one full
pitch of the double helix and periodicity is enforced in the program.
The population densities for the highest occupied molecular orbital
(HOMO) and the lowest unoccupied molecular orbital (LUMO) are plotted in
Fig.\ \ref{figHLperiodic}.  As seen from this figure, both the HOMO and
LUMO states exhibit very extended and periodic (Bloch-like) states
throughout the molecule. No localization is evident.

\subsection{Electronic states of sampled poly(dA)-poly(dT) DNA configurations}
\label{sec-short-dynamicDNA}

\begin{figure}[tb]

\begin{center}\includegraphics[%
  width=\figwidth]{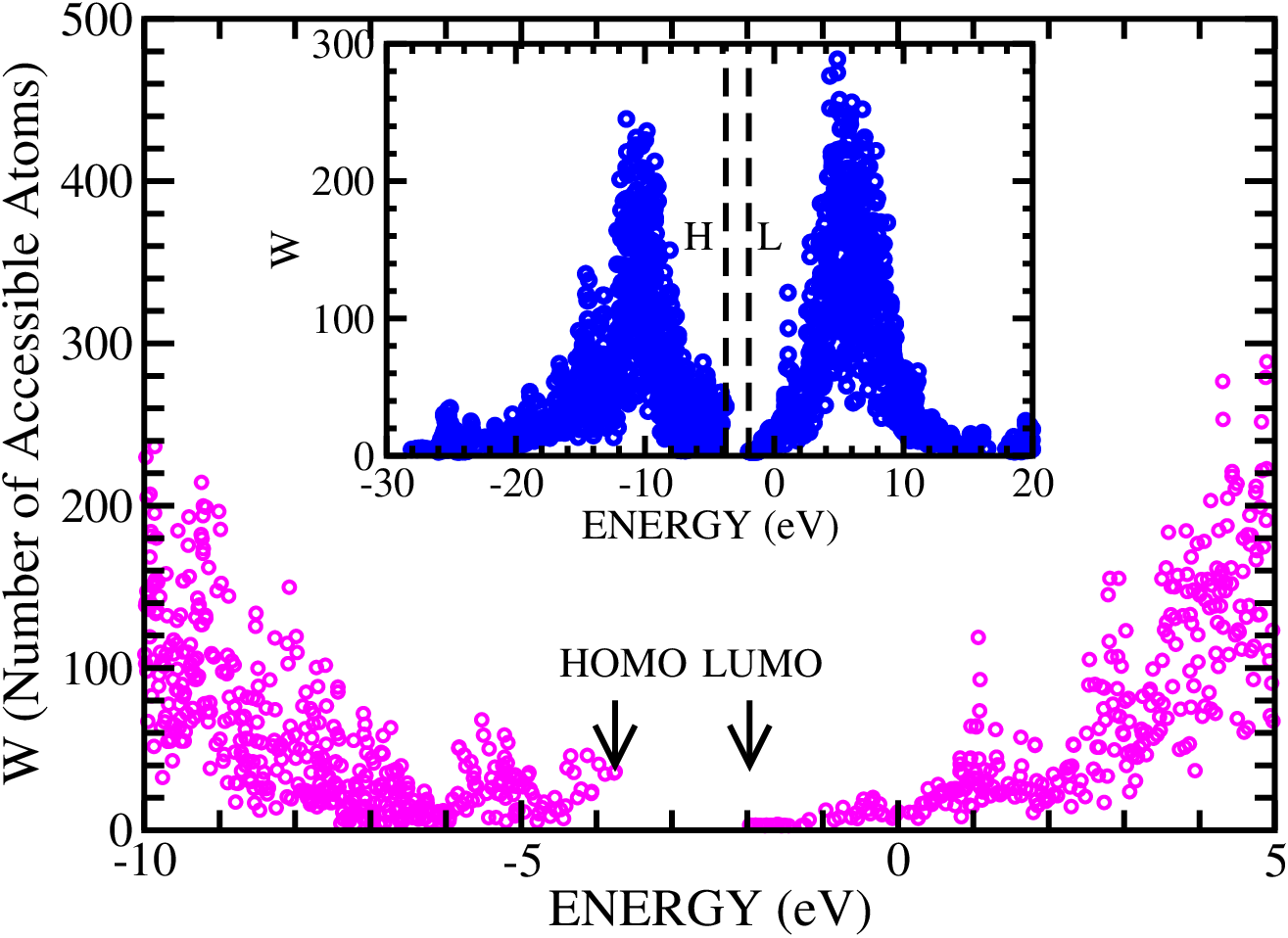}\end{center}
\caption{\label{figW_1}
  Number of accessible atoms, \protect$W(\nu)$, for each electronic
  state near the HOMO and LUMO levels. Inset shows number of accessible
  atoms for all levels. The system contains 10 basepairs of DNA (644
  atoms).  }

\end{figure}

We now consider results for a single configuration from the MD
simulation (labeled step 3001, the first coordinate set 0.5 ps after a
1.5 ns production simulation). Figure \ref{figW_1} shows the number of
accessible atoms, $W(\nu)$, for each electronic state at this time step
for the dehydrated structure. The $W(\nu)$ for the hydrated DNA
structure is shown in Fig. \ref{figW_2}. For both structures it is
important to note that, near the HOMO and LUMO, the number of accessible
atoms is quite small ($<30$), demonstrating a large degree of
localization for the wavefunctions. This localization extends over
several $e$V and is deeper than just the band tail states. States
further away from the HOMO and LUMO become considerably delocalised and
the number of accessible atoms is much larger. The number of accessible
atoms is also small for the lowest energy levels; these deep states
consist mainly of 2s levels of oxygen and nitrogen atoms.  For the
hydrated DNA molecule, the localized states near the HOMO are mainly due
to the surrounding water molecules. Just below these water-related
localized electronic states are the localized electronic states residing
on the DNA bases.  This may account for the smaller band gap of the
hydrated structure compared with the electronic structure of the
dehydrated DNA. Overall, the electronic structures for both the hydrated
and dehydrated DNA molecules show remarkable similarities.  These
results imply that the aquatic environment does not significantly alter
DNA's electronic structure. Therefore, we focus our studies on the
electronic structures of dehydrated DNA molecules.

\begin{figure}[tb]

\begin{center}\includegraphics[%
  width=\figwidth]{figW_2.eps}\end{center}
\caption{\label{figW_2}
  Number of accessible atoms, \protect$W(\nu)$, for each electronic
  state near the HOMO and LUMO levels. Inset shows number of accessible
  atoms for all levels. The system contains 10 basepairs of DNA (644
  atoms) and 350 water molecules.}

\end{figure}

\begin{figure}[tbhp]

\begin{center}\includegraphics[%
  width=0.70\columnwidth]{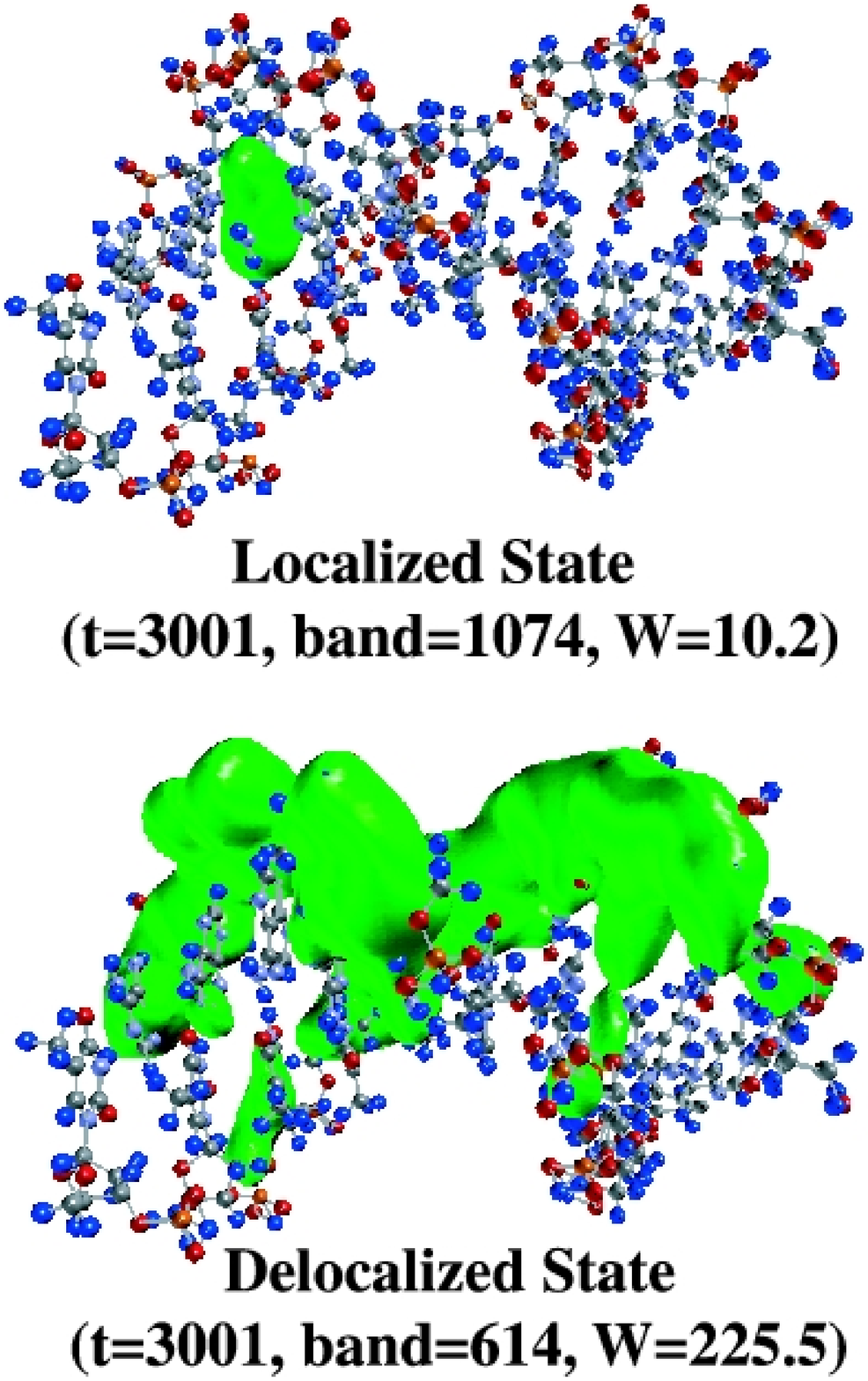}\end{center}
\caption{\label{figexample}
  Example of a localised and a delocalised state for two different
  states in poly(dA)-poly(dT) at time step 3001. For reference, the HOMO
  is band 1094.  }

\end{figure}

\begin{figure}[tb]

\begin{center}\includegraphics[%
  width=\figwidth]{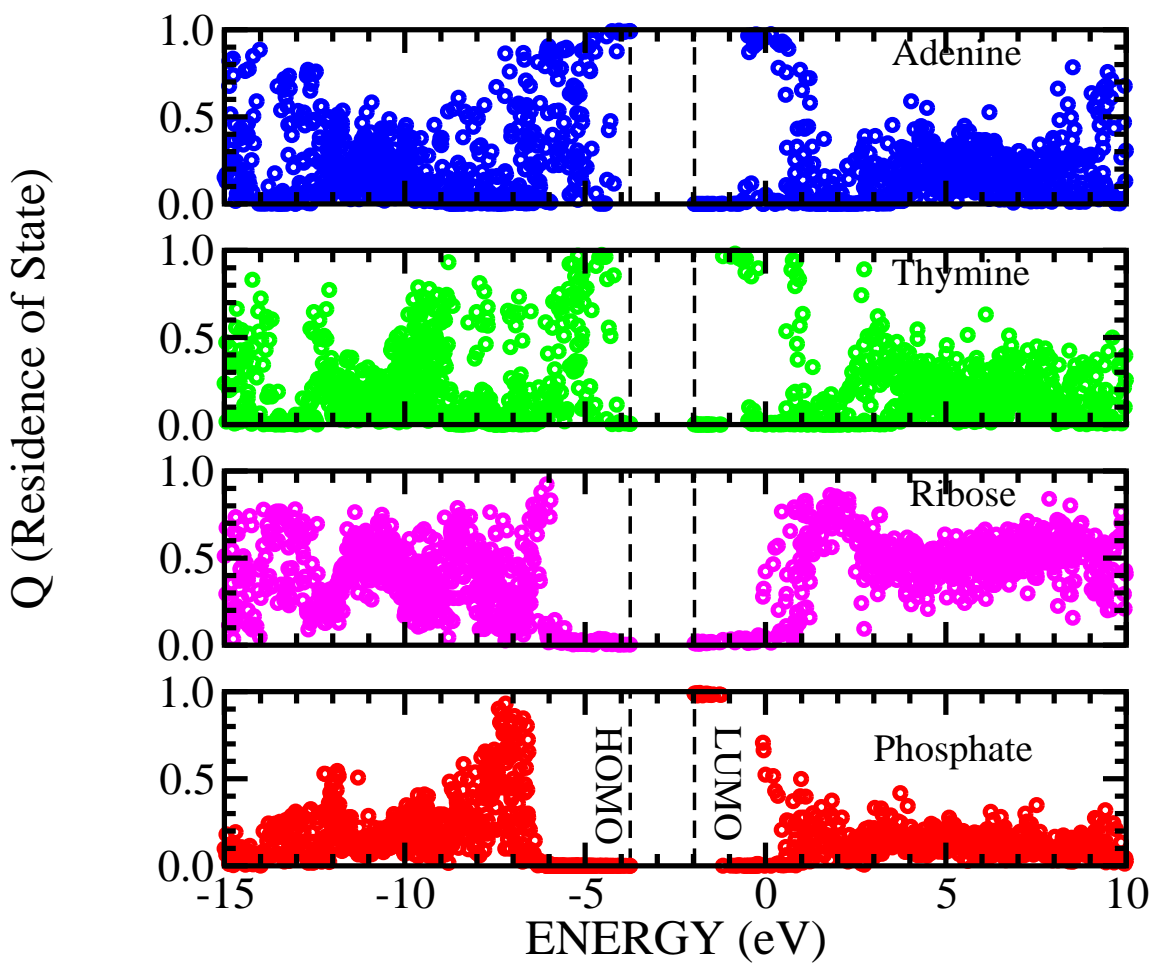}\end{center}
\caption{\label{figROS}
  Residence of state gives the location of the wavefunction for each
  energy state. States very near the HOMO level are located primarily on
  the adenine bases. For any given state, the sum of the four residences
  add to unity.  }

\end{figure}

The degree of localization for two example band states (1074 and 614 - larger
number implies higher eigenvalue) in the dehydrated DNA structures can be seen
in Fig.\ \ref{figexample} where population density plots of a localised and
delocalised state are shown.
As more configurations are analysed, we see consistently that the
number of accessible atoms for the energy levels near the HOMO primarily
consist of around 20 atoms. However, as a function of time, different
sets of atoms are involved. To determine where the localization occurs,
we compute a residence of each state according to the specific DNA
component - adenine base, thymine base, ribose backbone, or phosphate
group and determine where the high probability regions are located.
Further investigation indicates the residence localization for the
highly localised states near the HOMO are contained approximately
on single bases in the DNA molecule; adenine for states very near
the HOMO and thymine for states slightly lower in energy. This regional
population information for the HOMO on adenine is plotted in Fig.
\ref{figROS}. The more extended states ($\sim8$ $e$V to $\sim18$
$e$V below the HOMO) are found to reside throughout the various DNA
components.

\begin{figure}[tbhp]

\begin{center}\includegraphics[%
  width=0.90\columnwidth,
  angle=90]{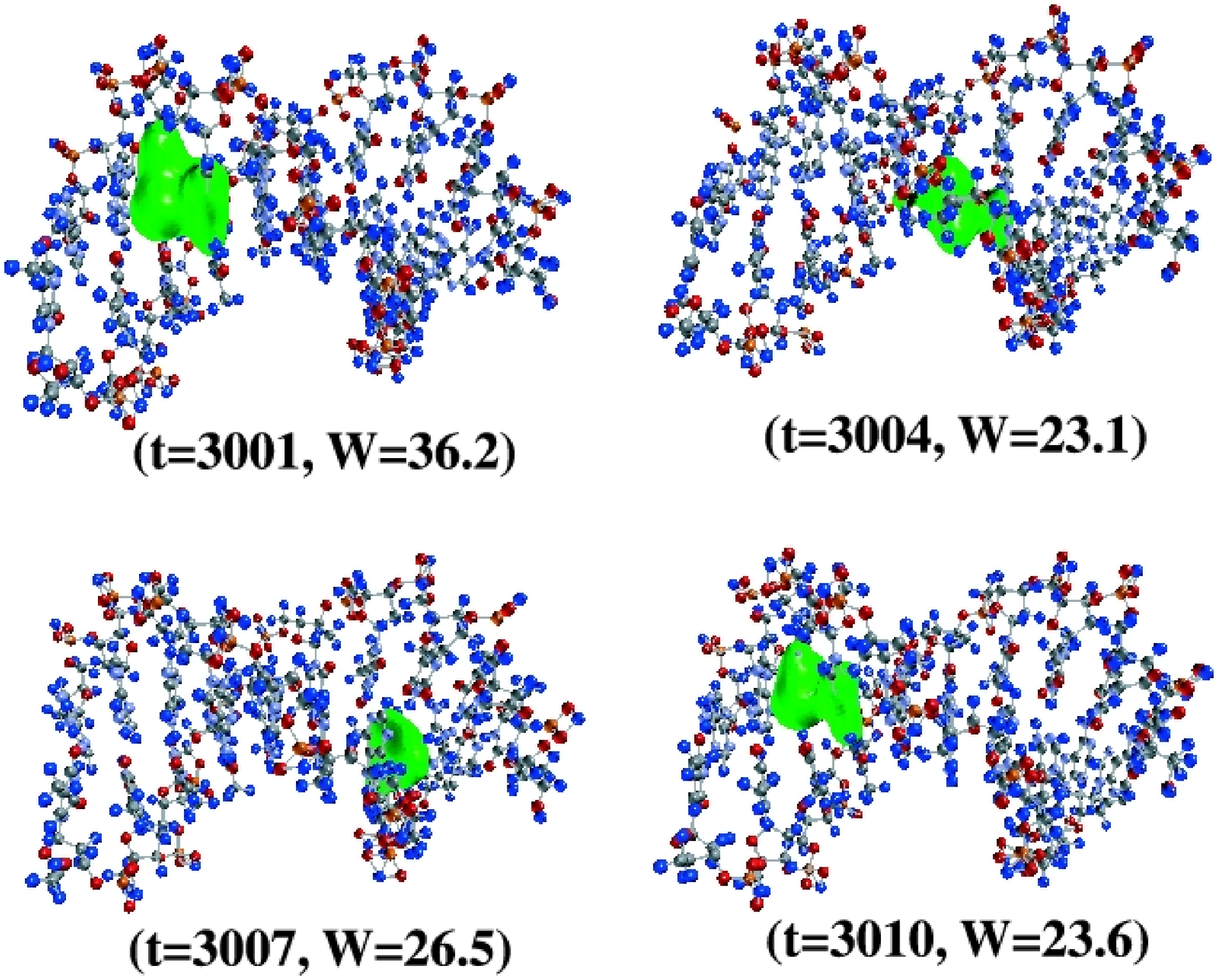}\end{center}
\caption{\label{figHOMOt1}
  Population density plots for the localised HOMO state as a function of
  time. The time between snapshots is 1.5 ps.  }

\end{figure}

\begin{figure}
\begin{center}\includegraphics[%
  width=\figwidth]{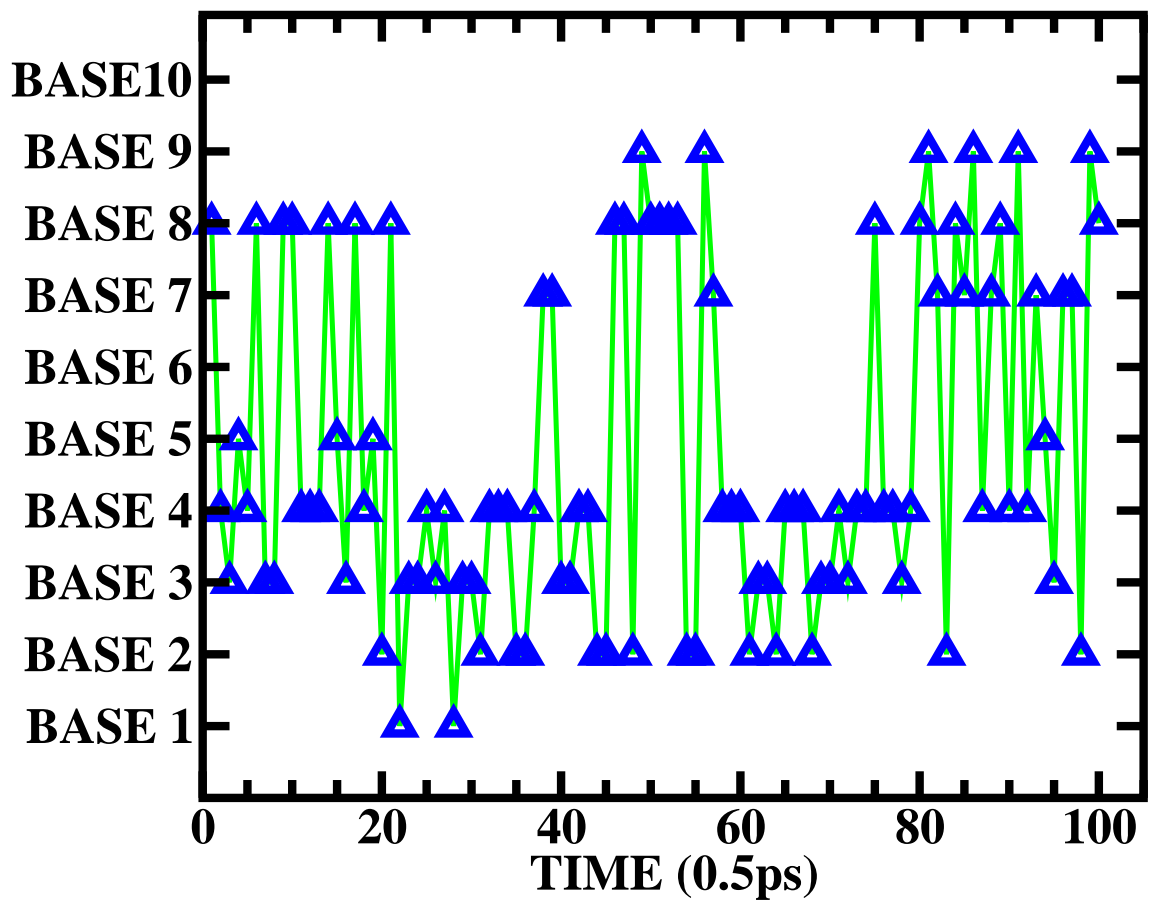}\end{center}
\caption{\label{figHOMOt2}
  Location of the HOMO as function of time. The ten bases are the ten
  adenine bases on one strand of the DNA. The HOMO
  is located only on adenine bases.
}
\end{figure}

As the simulation proceeds in time, the residence of the HOMO level
moves from base to base along the poly(dA)-poly(dT) system and large
jumps in sequence are possible over this 0.5 ps resolution time scale.
This fluctuating residency of the HOMO is visualised in Fig.\ \ref{figHOMOt1},
which shows population density plots for a series of snapshots at
different times (t=3001, 3004, 3007, and 3010). The separation between
these snapshots is 1.5 ps. Figure \ref{figHOMOt2} shows the location
of the HOMO for all 100 snapshots where the electronic structure was
calculated in this work. The population is localised on different
adenine bases as time progresses and appears to chaotically oscillate
between one end of the DNA molecule to the other. The HOMO level's
localization on one adenine base is traded for localization on another
adenine base through the dynamical simulation. Physically, this trading
ought to reflect concerted fluctuations assignable to off-diagonal
dynamical disorder in a regular homooligonucleotide duplex. Based
on these results, it is conceivable that electron (hole) transfer
will occur as two or more localised MO levels are dynamically trading
places. Moreover, this swapping may be gated by thermal fluctuations
of hydrated counterions, in accordance with the ion-gating transport
mechanism proposed in Ref.\ \cite{Schuster2001}.

Finally, it is of considerable interest to compare our above results to
the known literature data on this theme. Specifically, our findings are
in parallel with the most recently established dependence of electronic
coupling between DNA bases in the stack on DNA conformational states: a
diminuation of the coupling between the DNA purine bases due to the
pertinent conformational changes would 'arrest' the HOMO at one
particular base. Whereas, conformationally induced increases in the
above coupling ought to promote the 'HOMO trading' we revealed here. Our
results are also in accordance with the analogous approach put forth
most recently in Ref.\ \cite{Rosch2001} and in Ref.\ \cite{Orlandi2002}.
To be capable of formulating reasonable suggestions for
experimentalists, we would need more detailed calculations not only on
poly(dG)-poly(dC), but also on DNA with mixed base sequences.

\section{Effective tight-binding Hamiltonians for {\em long} DNA strands and complete sequences}
\label{sec-long}

In this section, we focus on whether DNA, when treated as a quantum wire
in the fully coherent low-temperature regime, is conducting or not. To
this end, we study and generalise a tight-binding model of DNA which has
been shown to reproduce experimental \cite{CunCPD02} as well as {\em
  ab-initio} results \cite{DavI04}. A main feature of the model is the
presence of sites which represent the sugar-phosphate backbone of DNA
but along which no electron transport is permissible. We emphasize that
the models is constructed to take into account the HOMO-LUMO gap
observed in the DFT-based studies in chapter \ref{sec-short} as well as
the observed absence of transport along the backbone.  We measure the
effectiveness of the electronic transport by the {\em localisation
  length} $\xi$, which roughly speaking parametrises whether an electron
is confined to a certain region $\xi$ of the DNA (insulating behaviour)
or can proceed across the full length $L$ ($\leq \xi$) of the DNA
molecule (metallic behaviour).

\subsection{The ladder model}
\label{sec-long-ladder}

A convenient tight binding model for DNA can be constructed as
follows: it has two central conduction channels in which
individual sites represent an individual base; these are
interconnected and further linked to upper and lower sites,
representing the backbone, but are \emph{not} interconnected along
the backbone. Every link between sites implies the presence of a
hopping amplitude. The Hamiltonian $H_L$ for this ladder-like
model is given by
\begin{eqnarray}
H_{L} &=& \sum_{i=1}^{L}
 \sum_{\tau=1,2}
    \left( t_{i,\tau}|i,\tau\rangle \langle i+1,\tau| +
    \varepsilon_{i,\tau} |i,\tau\rangle \langle i,\tau| \right)
 \nonumber \\
 & &  \mbox{}
+ \sum_{q=\uparrow,\downarrow}
    \left( t_i^q |i,\tau\rangle \langle i,q(\tau)|+
    \varepsilon_i^q|i,q\rangle \langle i,q| \right)
\nonumber \\
 & & + \sum_{i=1}^{L}
 t_{1,2}|i,1\rangle \langle i,2|
  \label{eq-ham2D}\label{eq-ladder}
\end{eqnarray}
where $t_{i,\tau}$ is the hopping amplitude between sites along each
branch $\tau=1$, $2$ and $\varepsilon_{i,\tau}$ is the corresponding
onsite potential energy. $t_i^q$ and and $\varepsilon_i^q$ give hopping
amplitudes and onsite energies at the backbone sites. Also,
$q(\tau)=\uparrow, \downarrow$ for $\tau=1, 2$, respectively. The
parameter $t_{12}$ represents the hopping between the two central
branches, i.e., perpendicular to the direction of conduction. Quantum
chemical calculations with semi-empirical wave function bases using the
SPARTAN package \cite{Spartan} results suggest that this value,
dominated by the wave function overlap across the hydrogen bonds, is
weak and so we choose $t_{12}= 1/10$.\footnote{Simulations with larger
  $t_{12}\sim 1/2$ give qualitatively similar results.} As we restrict our
attention here to pure DNA, we also set $\varepsilon_{i,\tau}=0$ for all
$i$ and $\tau$. Note that in this way, the energy gap has been made to
be symmetric about $E=0$. Hence when comparing with the results in
section \ref{sec-short}, a constant shift according to the neglected
ionisation potentials has to be added.

The model (\ref{eq-ladder}) clearly represents a dramatic
simplification of DNA. Nevertheless, in Ref.\ \cite{CunCPD02} it
had been shown that an even simpler model --- in which base-pairs
are combined into a single site --- when applied to an artificial
sequence of repeated GC base pairs, poly(dG)-poly(dC) DNA,
reproduces experimental data current-voltage measurements when
$t_{i}=0.37 e$V and $t_i^q=0.74 e$V are being used. This motivates
the above parametrisation of $t_i^q = 2 t_{i}$ and
$t_{i,\tau}\equiv 1$ for hopping between like (GC/GC, AT/AT)
pairs. Assuming that the wave function overlap between consecutive
bases along the DNA strand is weaker between unlike and
non-matching bases (AT/GC, TA/GC, etc.) we thus choose $1/2$.
Furthermore, since the energetic differences in the adiabatic
electron affinities of the bases are small \cite{WesLPS01}, we
choose $\varepsilon_{i}=0$ for all $i$. Due to the
non-connectedness of the backbone sites along the DNA strands, the
model (\ref{eq-ladder}) can be further simplified to yield a model
in which the backbone sites are incorporated into the electronic
structure of the DNA. The effective ladder model reads as
\begin{eqnarray}
\tilde{H}_{L} & = & \sum_{i=1}^{L}
  t_{1,2}|i,1\rangle \langle i,2| +
\sum_{\tau=1,2}
    t_{i,\tau}|i,\tau\rangle \langle i+1,\tau|
\nonumber \\
& & \mbox{ }  + \left[ \varepsilon_{i,\tau} -
 \frac{\left(t_{i}^{q(\tau)}\right)^{2}}{\varepsilon_{i}^{q(\tau)} - E}
\right] |i,\tau\rangle \langle i,\tau| + h.c. \quad .
\label{eq-ladder-effective}
\end{eqnarray}
Thus the backbone has been incorporated into an {\em
energy-dependent} onsite potential on the main DNA sites. This
effect is at the heart of the enhancement of localization lengths
due to increasing binary backbone disorder reported previously
\cite{KloRT05}.

\subsection{The numerical approach to localisation in a Hamiltonian tight-binding model}
\label{sec-long-localization}

There are several approaches suitable for studying the transport
properties of the model (\ref{eq-ladder}) and
these can be found in the literature on transport in solid state
devices, or, perhaps more appropriately, quantum wires. Since the
variation in the sequence of base pairs precludes a general solution, we
will use two methods well-known from the theory of disordered systems
\cite{RomS03}.

The first method is the iterative transfer-matrix method (TMM)
\cite{PicS81a,PicS81b,MacK83,KraM93,Mac94} which allows us in principle
to determine the localisation length $\xi$ of electronic states in
systems with cross sections $M=1$ \cite{CunCPD02} and $2$ (ladder) and
length $L \gg M$, where typically a few million sites are needed for $L$
to achieve reasonable accuracy for $\xi$. However, in the present
situation we are interested in finding $\xi$ also for viral DNA strands
of typically only a few ten thousand base-pair long sequences.  Thus in
order to restore the required precision, we have modified the
conventional TMM and now perform the TMM on a system of fixed length
$L_0$. This modification has been previously used
\cite{FraMPW95,RomS97b,NdaRS04} and may be summarised as follows:
After the usual forward calculation with a global transfer matrix ${\cal
  T}_{L_0}$, we add a backward calculation with transfer matrix ${\cal
  T}^{\rm b}_{L_0}$. This forward-backward-multiplication procedure is
repeated $K$ times. The effective total number of TMM multiplications is
$L_{\rm }=2KL_0$ and the global transfer-matrix is ${\tau}_{L_{\rm }} =
\left( {\cal T}^{\rm b}_{L_0} {\cal T}_{L_0}\right)^K$. It can be
diagonalised as for the standard TMM with $K\rightarrow \infty$ to give
${\tau}^{\dagger}_{L_{\rm }} {{\tau}_{L_{\rm }}} \rightarrow \exp[ {\rm
  diag}(4KL_0/\xi_{\tau})]$ with $\tau=1$ or $\tau= 1, 2$ for fishbone
and ladder model, respectively. The largest $\xi_{\tau}$ for all $\tau$
then corresponds to the localisation lengths of the electron on the DNA
strand and will be measured in units of the DNA base-pair spacing
($0.34$ nm).

The second method that we will use is the recursive Green function
approach pioneered by MacKinnon \cite{Mac80,Mac85}. It can be used to
calculate the dc and ac conductivity tensors and the density of states
(DOS) of a $d$-dimensional disordered system and has been adopted to
calculate all kinetic linear-transport coefficients such as
thermoelectric power, thermal conductivity, Peltier coefficient and
Lorentz number \cite{RomMV03}.

The main advantage of both methods is that they work reliably (i) for
short DNA strands ranging from 13 (DFT studies \cite{PabMCH00}) base
pairs up to 30 base pairs length which are being used in the nanoscopic
transport measurements \cite{DavI04} as well as (ii) for somewhat longer
DNA sequences as modelled in the electron transfer results and (iii)
even for complete DNA sequences which contain, e.g.\ for human
chromosomes up to 245 million base pairs \cite{AlbBLR94}.




\subsection{Long DNA sequences: $\lambda$-DNA, centromers and (super-)promoters}
\label{sec-long-DNA}

We shall use $2$ naturally occurring long DNA sequences (``strings").  (i)
$\lambda$-DNA \cite{lambda} is DNA from the bacteriophage virus. It has
a sequence of 48502 base pairs and is biologically very well
characterised. Its ratio $\alpha$ of like to un-like base-pairs is
$\alpha_{\lambda}=0.949$. (ii) centromeric DNA for chromosome 2 of yeast
has $813138$ base pairs \cite{cen2} and $\alpha_{{\rm centro.}}=0.955$.
This DNA is also rich in AT bases and has a high rate of repetitions
which should be favourable for electronic transport.

Another class of naturally existing DNA strands is provided by so-called
promoter sequences. We use a collection of 4986 is these which have been
assembled from the TRANSFAC database and cover a range of organisms such
as mouse, human, fly, and various viruses.  Promoter sequences are
biologically very interesting because they represent those places along
a DNA string where polymerase enzymes bind and start the copying process
that eventually leads to synthesis of proteins. On average, these
promoters consist of approximately $17$ base-pairs, much too short for a
valid localization length analysis by TMM. Therefore, we concatenate
them into a $86827$ base-pair long {\em super-promoter} with
$\alpha_{{\rm super-p.}}=0.921$. In order to obtain representative
results, $100$ such super-promoters have been constructed, representing
different random arrangements of the promoters, and the results
presented later will be averages. As usual, averages of $\xi$ are
computed by averaging the normally distributed $1/\xi$ values.

Occasionally, we show results for ``scrambled'' DNA. This is DNA with
the same number of A, T, C, G bases, but with their order randomised.
Clearly, such sequences contain the same set of electronic potentials
and hopping variations, but would perform quite differently in a
biological context. A comparison of their transport properties with
those from the original sequence thus allows to measure how important
the exact fidelity of a sequence is. On average, we find for these
sequences $\alpha_{\lambda/{\rm S}}=0.899$, $\alpha_{{\rm centro./S}}=
0.9951$ and $\alpha_{{\rm super-p./S}}= 0.901$.

A convenient choice of artificial DNA strand is a simple, $100000$
base-pair long {\em random} sequence of the four bases, random-ATGC DNA,
which we construct with equal probability for all 4 bases ($\alpha_{{\rm
    random}}=0.901$). We shall also `promote' these random DNA strings
by inserting all $4086$ promoter sequences at random positions in the
random-ATGC DNA ($\alpha_{{\rm random/P}}= 0.910$).

\subsection{Results for localization lengths}
\label{sec-long-results}

We have computed the energy dependence of the localization lengths
for all sequences of section \ref{sec-long-DNA}. In addition,
$\lambda$-DNA, centromeric DNA and the super-promoter DNA where
also scrambled $100$ times and the localization length of each
resulting sequence measured and the appropriate average
constructed. Also, we constructed $100$ promoted random-ATGC DNA
sequences.
\begin{figure}[tb]
  \centering
  \includegraphics[width=\figwidth]{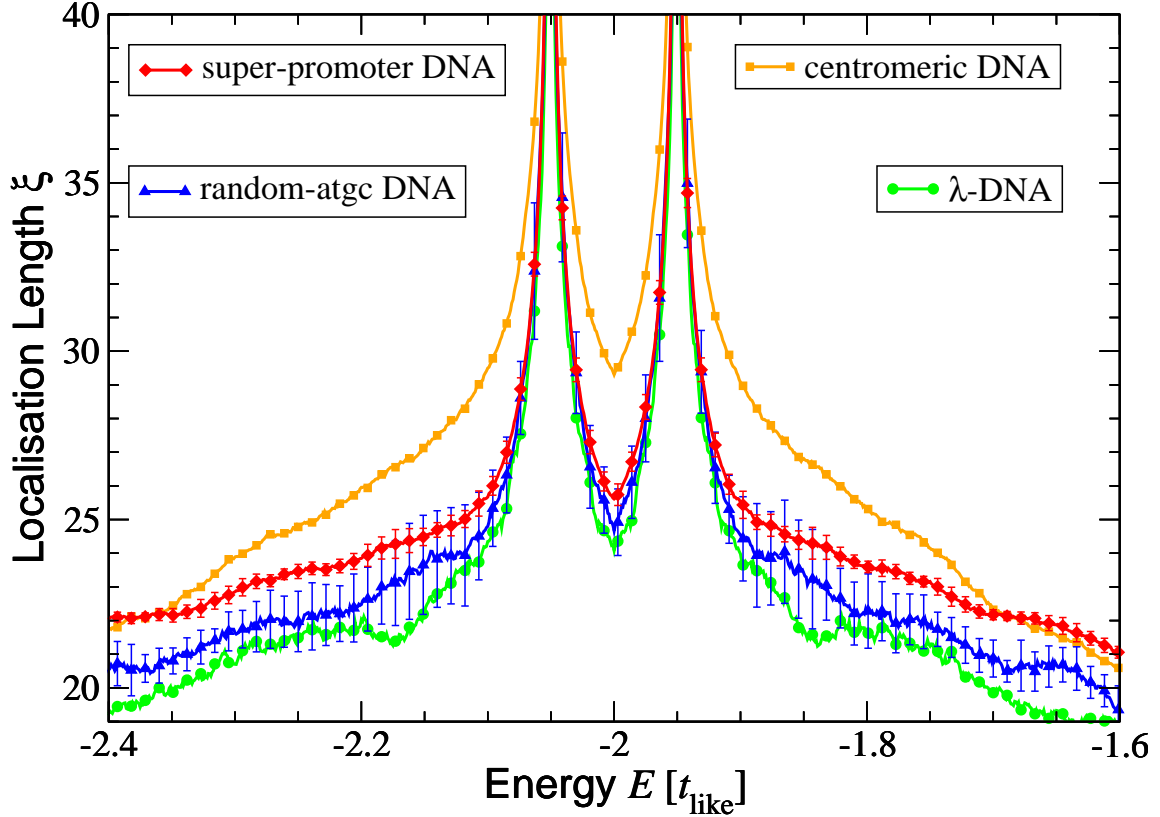}
  \caption{\label{fig-DD-Loc_Energy-origDNA}
    Localization lengths $\xi$ versus Fermi energy $E$ for various clean
    DNA strands. Only every 10th symbol is shown. Error bars reflect the
    standard deviation after sampling the different sequences for
    random-ATGC and promoted DNA. The energy is measured in unit of
    hopping energy between like base pairs, i.e., $t_{\rm
      like}=t_{i}=0.37 e$V.}
\end{figure}
As shown previously \cite{KloRT05}, the energy dependence of $\xi$
reflects the backbone-induced two-band structure. The obtained $\xi(E)$
values for the lower band are shown in Fig.\
\ref{fig-DD-Loc_Energy-origDNA}. In the absence of any onsite-disorder,
we find two prominent peaks separated by $t_{1,2}$ and $\xi(E)=\xi(-E)$.
We also see that $\lambda$-DNA has roughly the same $\xi(E)$ dependence
as random-ATGC-DNA. The super-promoter has larger $\xi$ values compared
to random-atcg- and $\lambda$-DNA. Most surprisingly, centromeric DNA
--- the longest investigated DNA sequence --- has a much larger
localization length than all other DNA sequences. The order of
like-to-unlike pair-ratios is $\alpha_{\rm centro.}= 0.955 >
\alpha_{\lambda}= 0.949 > \alpha_{\rm super-p.}= 0.921 > \alpha_{\rm
  random} = 0.901$ and one might expect that transport is favoured in
sequences with large $\alpha$. From Fig.\
\ref{fig-DD-Loc_Energy-origDNA}, it is clear that this is not the case,
$\lambda$-DNA has the smallest localization lengths, but the second
largest $\alpha$.

\begin{figure}[tb]
  \centering
  \includegraphics[width=\figwidth]{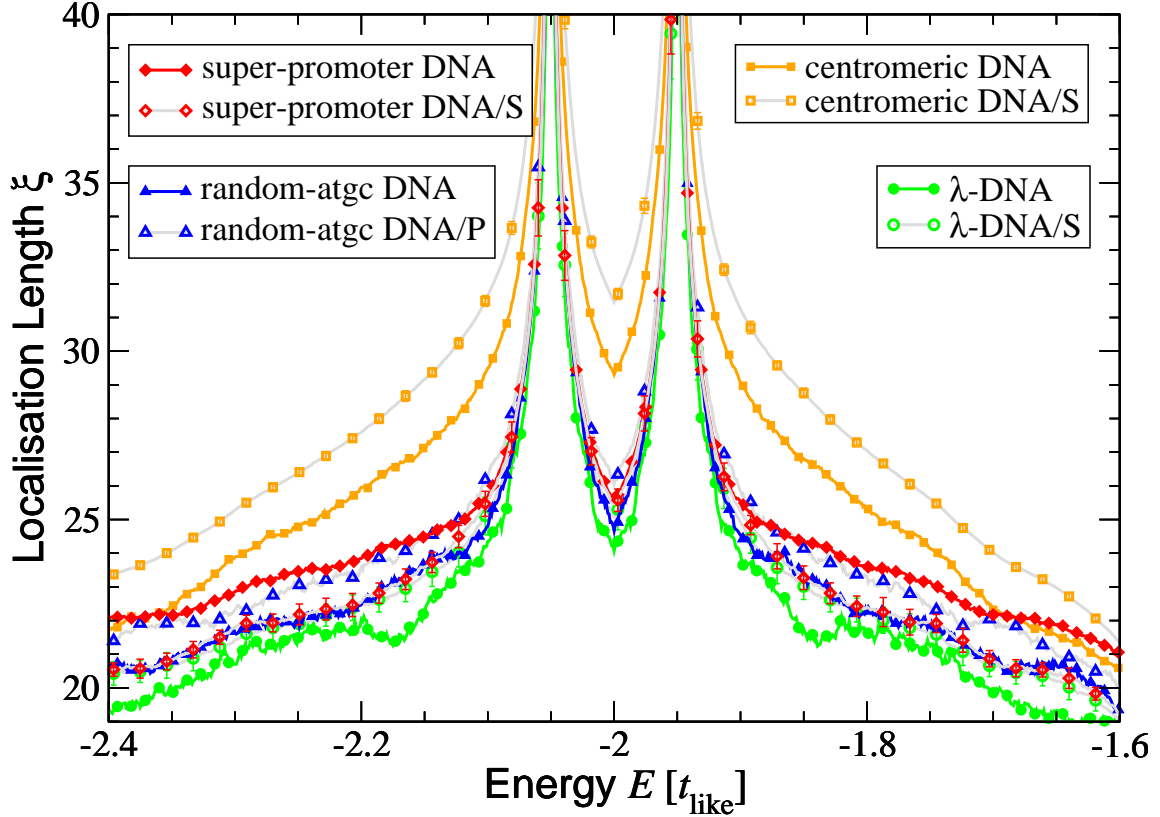}
  \caption{\label{fig-DD-Loc_Energy-PS}
    Localization lengths $\xi$ versus Fermi energy $E$ for various clean
    DNA (solid symbols as in Fig.\ \ref{fig-DD-Loc_Energy-origDNA},
    error bars not shown for clarity), scrambled DNA (DNA/S, (open
    $\diamond$, $\Box$, $\circ$) and promoted DNA (DNA/R, open
    $\triangle$) strands. Only every 10th (20th) symbol is shown for
    clean (scrambled/promoted) DNA. Error bars reflect the standard
    deviation after sampling the different sequences for random-ATGC,
    scrambled and promoted DNA.}
\end{figure}

In Fig.\ \ref{fig-DD-Loc_Energy-PS}, we add results for scrambled and
promoted DNA.
We find that promoting a given DNA sequence leads to small increases in
localization length $\xi$ for random DNA, whereas scrambling can lead to
increase (centrometric and $\lambda$-DNA) as well as decrease
(super-promoter). These results suggest that the promoters have a
tendency towards larger localization lengths and thus enhanced
transport.

\subsection{Promoter sequences and E.\ coli binding sites}
\label{sec-long-promoters}

Let us now turn our attention to the transport properties of individual
promoters rather than the artificially constructed super-promoters.
Since their average lengths is 17 base-pairs and thus comparable to the
localization lengths measured in the longer sequences, we can no longer
use the TMM, but need to employ the RGFM mentioned in Section
\ref{sec-long-localization}. While this method is capable of computing
all thermoelectric transport coefficients, we shall restrict ourselves
to presenting results for the conductance here.

\begin{figure}[tb]
  \centering
  \includegraphics[width=\figwidth]{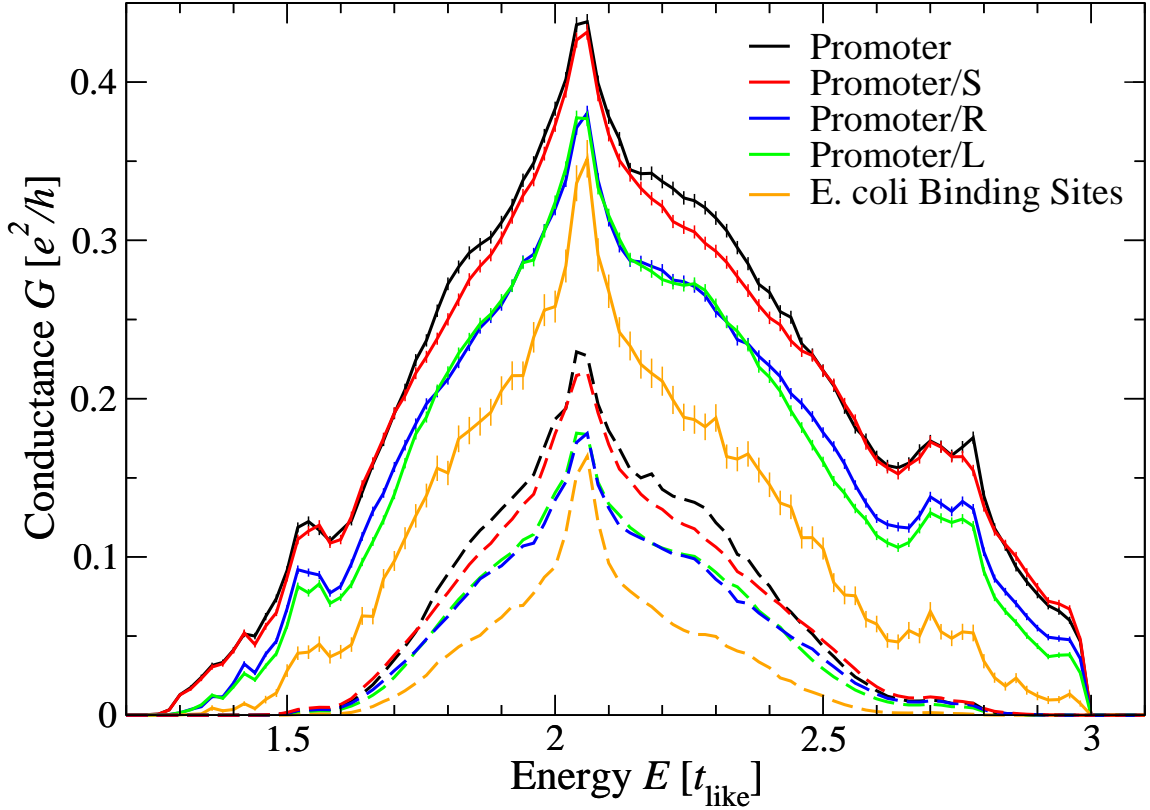}
  \caption{\label{fig-promo-COND_Energy}
    Energy dependence of the conductance $G$ for promoters, scrambled
    promoters (/S), random promoters (/R) and $\lambda$-promoters (/L).
    Solid lines denote the arithmetic, dashed lines the typical average
    of $G$. The error bars denote standard deviation obtained from the
    $4986$ different promoters considered in each category (original and
    /S, /R, /L). These are not repeated for the typical averages for
    clarity.}
\end{figure}

In Fig.\ \ref{fig-promo-COND_Energy}, we show results for averaged
conductance in the upper band; both arithmetic and typical conductance
have been calculated.
We first note that the double-peak structure of Fig.\
\ref{fig-DD-Loc_Energy-origDNA} has vanished and only a single peak
remains. This is because our results have been computed with
perfectly-conducting leads attached to both ends of the DNA strands.
This is close to the experimental situation, but the purely off-diagonal
disorder in the DNA model is now masked by the ordered leads.
Next, we observe that the promoters and their scrambled copies have
larger conductances than random- and $\lambda$-promoters.
$\lambda$-promoters has been constructed by cutting sequences with the
same lengths as the true promoters out of $\lambda$-DNA at randomly
selected positions along the DNA. Since $\alpha$ for random and
$\lambda$-DNA is different, this allows us to check whether it is the
order of base pairs or the value of $\alpha$ which dominated the value
of $G$. Since $\alpha_{\rm promoter}= 0.928 < \alpha_{\lambda}=0.955$,
but $G_{\rm promoter} > G_{\lambda}$, it appears that as before the
transport properties are not simply large if $\alpha$ is large. This
suggests that it is indeed the fidelity of the sequence which is also
important.

\begin{figure}[tb]
  \centering
  \includegraphics[width=\figwidth]{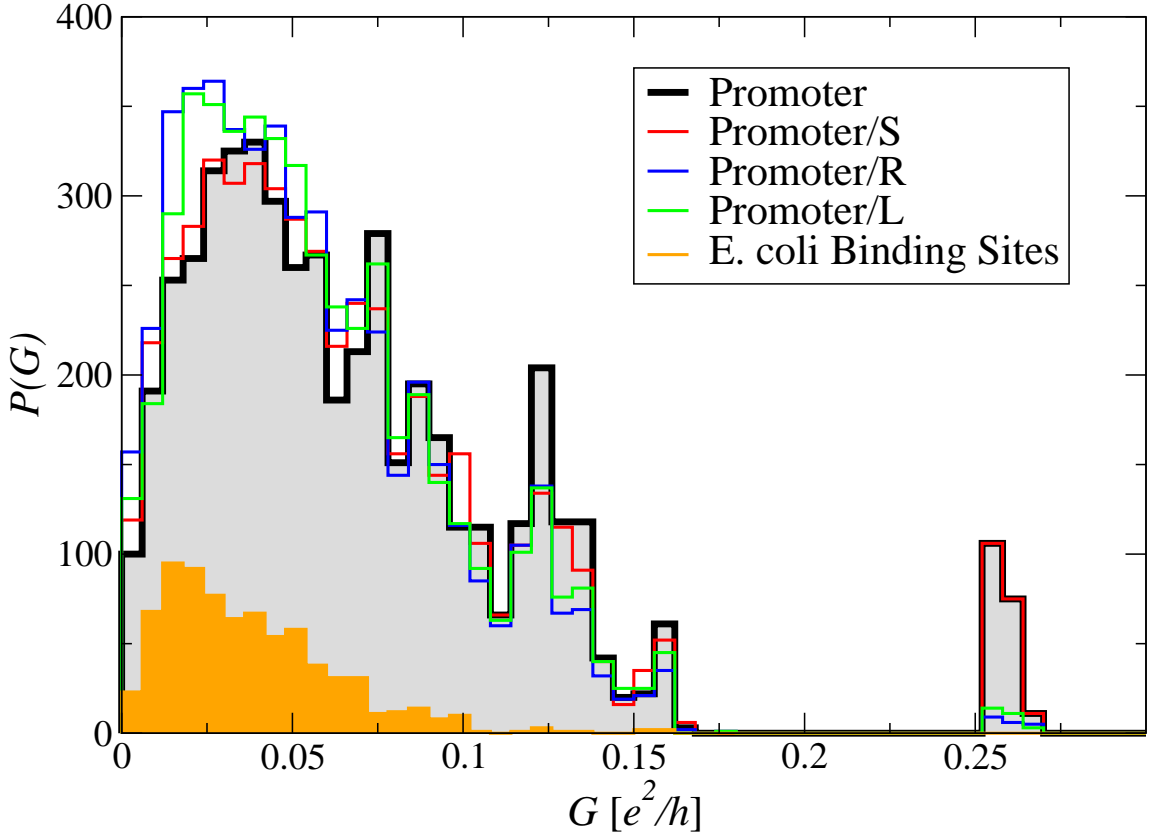}
  \caption{\label{fig-promo-histo-COND}
    Distribution function $P(G)$ for the conductances averaged over the
    energy range $[-5,5]$. Only promoter and E.\ coli results have been
    shaded.}
\end{figure}

Typical and arithmetic averages share similar characteristics when
comparing different sequences as shown in Fig.\
\ref{fig-promo-COND_Energy}. However, the typical values are
systematically smaller than their arithmetic counterparts. We therefore
expect the distributions to be highly non-Gaussian and in Fig.\
\ref{fig-promo-histo-COND} we see that this is indeed the case.
We first note that both the original promoter as well as their scrambled
version (/S) appear to have a slightly larger weight at $G > 0.05$ whereas
both random and $\lambda$-DNA are peaked at $G\sim 0.025$.
In addition, we find that there is a peak in the conductance
distribution $P(G)$ at $G \sim 0.26$. This peak is most pronounced for
the original promoter and their scrambled cousins, but much smaller for
the artificial random- and $\lambda$-promoter.

In Figs.\ \ref{fig-promo-COND_Energy} and \ref{fig-promo-histo-COND}, we
have also included results for computationally inferred 802 E.\ coli
bindings sites \cite{RobMC98}. Sequence-specific DNA-binding proteins
perform a variety of roles in the cell, including transcriptional
regulation. Our results show that the total conductance of these
sequences is smaller than for promoters. However, their average length
is $\sim 25$ so that the average {\em conductivity} is in fact larger
when compared to promoters. This might be important in a biological
context where one could envisage proteins to identify their binding
sites differences on local conductivities.

\section{Summary}
\label{sec-summary}

The results presented in this chapter are preliminary results but
indicate a marked difference in the nature of the electronic HOMO-LUMO
states for the periodic and aperiodic structures of duplex DNA. These
results indicate that the HOMO-LUMO states for the periodic structure
are quite extended as would be expected for Bloch-like states while the
HOMO-LUMO states for the aperiodic structure demonstrates more
localization. The concept of static localization in short DNA has previously
been considered by Ladik \cite{Bakhshi1986,Ladik2001}, and our results show that
such a localization in our structure for aperiodic poly(dA)-poly(dT) DNA
reaches far deeper in energy than just the band tail states. The
localization phenomenon observed in the DNA double helix is the
so-called Anderson localization which attributes to the
\emph{off-diagonal} disorder. This disorder results from dynamical
variations in DNA intramolecular interactions and coupling of DNA with
its environment.
Turning our attention to longer DNA sequences, we next used this insight
by modelling DNA as an off-diagonally disordered Anderson chain.
However, in addition and contradistinction to previous studies using
Anderson-type models, we include the sugar-phosphate backbone
explicitly and by doing so retain the essential semi-conducting
structure as observed in some experiments. Our results for the
localization lengths suggest extended states even in non-periodic DNA up
to $\sim 20$ base-pairs distances. This is roughly consistent with the
previous results. Next, we study how transport properties differ between
sequences and find that promoter sequences seem to have a tendency
towards larger localization length, i.e.\ enhanced transport. This might
point towards the importance of an electronic mechanism in the initial
stages of DNA polymerase.
Our results warrant further investigation, as the role of the dynamical
localization and the sequence dependence may very well suggest an
important mechanism of charge transport along the DNA molecule.

\subsection*{Acknowledgements}

RAR gratefully acknowledges discussions with A. Rodriguez and M.S. Turner.
JPL acknowledges discussions with D. Drabold, O. F. Sankey and T. Cheatham
who prepared MD simulations. While the chapter has been prepared jointly,
calculations in Sec. \ref{sec-short} were done by J. P. Lewis, H. Wang and
R. Marsh, whereas Sec. \ref{sec-long} is based on results of R. A. R\"{o}mer.


\end{document}